\documentclass[a4paper,11pt]{article}
\usepackage{pos}

\title{Chiral Symmetry and Large Magnetic Fields}

\author[a,b,\dag]{Prabal Adhikari}
\author*[c,d,e,\ddag]{Brian C.~Tiburzi}

\note[\dag]{Research supported in part by the 
U.S.~Department of Energy, 
Office of Science, 
Office of Nuclear Physics and Quantum Horizons Program under Award Number 
DE-SC0024385, 
and by the KITP through the U.S.~National Science Foundation under Grant 
No.~NSF PHY-1748958.}

\note[\ddag]{Research supported in part by funds provided by the Excellence Network Initiative at the Forschungszentrum J\"ulich under FK EXNET-$01$-$07$.}

\affiliation[a]{Physics Department, 
        Faculty of Natural Sciences and Mathematics,
        St.~Olaf College,\\
        Northfield, 
        MN 55057, USA}

\affiliation[b]{Kavli Institute for Theoretical Physics, 
	University of California,\\
	Santa Barbara, 
	CA 93106, 
	USA}
	
\affiliation[c]{Department of Physics,
        The City College of New York,\\
        New York,
        NY 10031, USA}	
        
\affiliation[d]{Graduate School and University Center,
        The City University of New York,\\
        New York, 
        NY 10016, 
        USA}        

\affiliation[e]{Institute for Advanced Simulation (IAS-4), 
      	Forschungszentrum J\"ulich,\\
	52428 J\"ulich, Germany
	}

\emailAdd{adhika1@stolaf.edu}
\emailAdd{btiburzi@ccny.cuny.edu}

\abstract{
Large magnetic fields exist in magnetars and are produced in off-central heavy-ion collisions. 
For the latter,  
field strengths are estimated to be comparable to strong interaction scales. 
This fact has motivated many studies of QCD physics in large magnetic fields, 
ranging from various model studies to lattice QCD computations. 
We provide a selective overview of results stemming from chiral perturbation theory. 
These results are based solely on the pattern of spontaneous and explicit symmetry 
breaking of QCD in a magnetic field; 
accordingly, 
they constitute low-energy theorems that must be satisfied in any approach. 
A few discrepancies with models and tension with lattice data are highlighted. 
}

\FullConference{%
	 42nd International Conference on High Energy Physics (ICHEP2024)\\
  	18-24 July 2024\\
  	Prague, Czech Republic\\
}

\begin{document}
\maketitle

\section{Motivation and Overview}

The study of QCD in large magnetic fields is motivated by extreme hadronic environments. 
Recently, 
a record-breaking surface magnetization of a magnetar was observed%
~\cite{Kong:2022cbk}. 
While the field 
$1.6 \times 10^{13} \, \texttt{Gauss}$
($e B \approx 0.3 \, \texttt{MeV}^2$)
falls short of QCD scales, 
field strengths inside these highly magnetized neutron stars are conjectured to be considerably 
greater than those that persist on the surface. 
Back on Earth, 
off-central heavy-ion collisions create the largest known magnetic fields. 
Microscopic transport models of the colliding ions estimate that magnetic fields are comparable to QCD scales, 
for example 
Ref.~\cite{Skokov:2009qp}. 
For heavy-ion collisions at RHIC, 
fields are estimated at 
$e B \sim m_\pi^2$, 
while those at LHC energies have fields upwards of
$e B \sim 15 m_\pi^2$. 
Quite recently, 
moreover, 
there has been experimental progress in probing the imprints of electromagnetic fields on particles emitted in heavy-ion collisions%
~\cite{STAR:2023jdd}. 
With these sizes, 
one must understand how strong interactions are modified in magnetic fields.

To address QCD in large magnetic fields, 
there are several approaches. 
Models are by far the most adaptable and can provide considerable intuition. 
Whether they are able to capture the correct non-perturbative physics of QCD, 
however,  
constitutes a perpetual challenge.%
\footnote{
We will compare chiral perturbation theory results with those obtained using the Nambu--Jona-Lasinio 
(NJL)
model, 
which incorporates some features of QCD.   
While we find discrepancies with the requirements of low-energy QCD, 
we note that other models are in far worse shape. 
} 
For magnetic fields, 
lattice gauge theory computations fortunately do not exhibit a sign problem, 
and technology developed to simulate QCD can be carried over for QCD in external magnetic fields
(albeit with considerable added computational cost for fully dynamical simulations). 
Additionally, 
there are constraints that arise from chiral perturbation theory
(ChPT). 
Such constraints are based solely on the pattern of spontaneous and explicit symmetry breaking 
of QCD, 
which can be appropriately amended to include external fields. 
These constraints essentially constitute low-energy theorems for QCD with magnetic fields, 
and are our primary concern throughout. 
While the maximum magnetic fields estimated for heavy-ion collisions at the LHC are well beyond the 
reach of ChPT, 
the low-energy theorems serve to anchor observables.
Only when an approach passes the stringent tests afforded by ChPT
will one gain confidence to extend such computations to larger field strengths.

\section{ChPT in a Magnetic Field}

In discussing ChPT, 
we focus on the systematic inclusion of the magnetic field, 
and then pursue a few applications.  
For simplicity, 
the framework is discussed strictly at zero temperature, 
with only zero-temperature results given below.

ChPT is the low-energy effective field theory of QCD 
based on the pattern of spontaneous symmetry breaking, 
with explicit symmetry breaking included perturbatively. 
For two massless light-quark flavors, 
QCD exhibits an 
$SU(2)_L \times SU(2)_R$
symmetry that is spontaneously broken to 
$SU(2)_V$
by the formation of the chiral condensate. 
Pions are the pseudo-Goldstone modes that emerge, 
but they are not massless due to explicit chiral symmetry breaking introduced by the light-quark masses.
With the addition of an external magnetic field, 
the chiral symmetry of massless QCD is reduced to 
$U(1)_L \times U(1)_R \subset SU(2)_L \times SU(2)_R$,
due to the difference in quark electric charges. 
This chiral symmetry is spontaneously broken to the vector subgroup
$U(1)_V \subset SU(2)_V$
by the formation of the chiral condensate. 
Spacetime symmetries are also altered by the inclusion of a static magnetic field. 
The Lorentz group 
$SO(3,1)$
is reduced to 
$SO(2) \times SO(1,1)$
due to the direction introduced by the magnetic field. 
Transverse to the magnetic field, 
one has planar rotational invariance;
longitudinal to the magnetic field, 
one has Lorentz boosts.

ChPT is formulated in terms of the coset field 
$U = \exp \left( i \vec{\pi} \cdot \vec{\tau} / F \right)$
containing the pseudo-Goldstone pion fields
$\vec{\pi}$ 
that emerge in the absence of explicit symmetry breaking introduced by
the quark masses and electric charges. 
The leading-order 
(LO)
chiral Lagrangian is that of the non-linear sigma model with symmetry breaking terms%
~\cite{Gasser:1983yg}
\begin{equation}
\mathcal{L}_2 
= 
\frac{F^2}{4} 
\big\langle 
D_\mu U^\dagger D^\mu U +  U^\dagger \chi + \chi^\dagger U
\big\rangle
\label{eq:L2}
,\end{equation}
where the quark masses appear in 
$\chi$,
and the external magnetic field appears in the covariant derivative
$D_\mu$. 
There are infinitely many higher-dimensional operators needed to renormalize this theory; 
however, 
as a low-energy effective theory, 
power counting is used to renormalize it with a finite set of operators required order-by-order in the chiral expansion. 
The systematic expansion is organized in terms of momentum and symmetry breaking parameters, 
for which 
$p^2 \sim m_\pi^2 \sim e B \ll (4 \pi F)^2 \approx 1.1 \, \texttt{GeV}^2$,
where 
$p$
represents a good component of the pion's momentum. 
While the magnetic field must be perturbatively small compared to the scale of ChPT, 
the field can still be large in physical terms because
$e B / m_\pi^2 \sim 1$. 
Expanding the LO Lagrangian to lowest order in the pion fields, 
one sees that the pion propagator requires the inclusion of all number of insertions of the magnetic field. 
This requires one to solve the underlying quantum mechanical problem of a pion in a magnetic field, 
however, 
closed-form expressions exist for the case of a uniform magnetic field, 
for example. 
We restrict our attention in what follows to the uniform-field problem. 
Summing insertions of the magnetic field results in Landau levels for the charged pion, 
while the neutral pion is unaffected at LO due to charge neutrality.

\section{Magnetic Masses of Pions}

While 
$\pi^0$
is unaffected at LO, 
the same is not true at next-to-leading order
(NLO). 
The Lagrangian 
Eq.~\eqref{eq:L2}
contains four-pion interaction terms
that renormalize the pion masses. 
The neutral pion receives loop corrections from both neutral- and charged-pion tadpoles, 
where the latter depends on the magnetic field. 
In ChPT, 
the energy of 
$\pi^0$ 
in a magnetic field
(its magnetic mass)
takes the form
\begin{equation} 
m_{\pi^0}^2(B)
=
m_\pi^2 
\left[ 
1 
+ 
\frac{eB}{(4 \pi F_\pi)^2}
\mathcal{I} \left( \frac{m_\pi^2}{eB} \right)
\right]
\label{eq:mpi0}
,\end{equation}
at NLO%
~\cite{Tiburzi:2008ma}. 
The factor 
$e B / (4 \pi F_\pi)^2$
is indicative of chiral corrections, 
while the non-analytic function 
\begin{equation}
\mathcal{I}(z)
=
2 \log \Gamma\left(\frac{1+z}{2}\right) 
+ z \left( 1 - \log \frac{z}{2} \right) - \log 2 \pi
\label{eq:I}
,\end{equation}
accounts for the effect of Landau levels from the charged-pion tadpole loop. 
Due to this function, 
one has
$m_{\pi^0}(B) < m_\pi$,
when 
$B \neq 0$. 
To get a sense of this result, 
one can expand about the weak-field limit 
$e B \ll m_\pi^2$, 
for which the magnetic mass becomes
\begin{equation}
m_{\pi^0}(B)
= 
m_\pi - \frac{1}{2} \beta^{\pi^0}_M B^2 + \mathcal{O}(B^4)
\quad 
\Rightarrow
\quad
\beta^{\pi^0}_M
=
\frac{e^2}{6 m_\pi (4 \pi F_\pi)^2}
,\end{equation}
where the coefficient at second-order in the magnetic field is the magnetic polarizability, 
and has been determined using
Eq.~\eqref{eq:mpi0}. 
Results for the magnetic mass of 
$\pi^0$ 
are obtained using the NJL model, 
for example in 
Ref.~\cite{Coppola:2018vkw}, 
and one can compare the behavior with that demanded by ChPT. 
To make the comparison succinct, 
we expand the NJL results to obtain the 
$\pi^0$
magnetic polarizability.
We find
\begin{equation}
\left( \beta_M^{\pi^0} \right)_\text{NJL}
\Big/ \beta_M^{\pi^0}
=
\left( \frac{F_\pi}{M} \right)^2 \frac{20/9}{- I_2(M,\Lambda)}
+ 
\mathcal{O} \left( \frac{m_\pi^2}{4 M^2} \right)
\approx 
\frac{1}{3}
,\end{equation} 
where 
$M$
is a constituent quark mass, 
and 
$\Lambda$
is an ultraviolet cutoff used to regulate the model in a renormalization group \emph{non}-invariant way. 
These parameters are not present in QCD, 
and we use the Set $1$ values and the loop function
$I_2$
provided in%
~\cite{Coppola:2018vkw}
for the numerical estimate.
The polarizability is in qualitative agreement:
it has the correct sign, 
but is otherwise too small.

For the charged pion, 
one can perform a similar analysis of its magnetic mass
$m_{\pi^-}(B)$, 
defined according to the NLO dispersion relation
$E_{\pi^-}^2
= 
m_{\pi^-}^2(B) + (2 n + 1) e B + p_\parallel^2$.
The magnetic mass 
$m_{\pi^+}(B) = m_{\pi^-}(B)$, 
due to charge conjugation invariance. 
Due to an accident, 
the NLO charged-pion mass in ChPT receives net loop contributions only from 
$\pi^0$. 
There is, 
however, 
an additional contribution from the tree-level insertion of the NLO operators%
~\cite{Gasser:1983yg}
\begin{equation}
\mathcal{L}_4 
\subset
l_5 
\big\langle 
U^\dagger \hat{R}^{\mu \nu} U \hat{L}_{\mu \nu} 
\big\rangle
+
l_6 \, \tfrac{i}{2} 
\big\langle 
D^\mu U^\dagger D^\nu U \hat{L}_{\mu \nu} 
+
D^\mu U D^\nu U^\dagger \hat{R}_{\mu \nu}
\big\rangle
\label{eq:L4}
,\end{equation}
which depend on trace-subtracted, 
right- and left-handed 
external field-strength tensors
$\hat{R}_{\mu \nu}$
and
$\hat{L}_{\mu \nu}$, 
respectively. 
With the tree-level contributions, 
the charged-pion magnetic mass at NLO in ChPT is
\begin{equation}
m_{\pi^-}^2(B) = m_\pi^2 + \overline{\ell} \left( \frac{e B}{4 \pi F_\pi} \right)^2
\quad
\Rightarrow
\quad
\beta_M^{\pi^-}
=
- \frac{e^2 \, \overline{\ell}}{m_\pi (4 \pi F_\pi)^2}
\label{eq:mpi-}
,\end{equation}
where the corresponding magnetic polarizability is indicated. 
The parameter 
$\overline{\ell} = \frac{1}{3} (\overline{\,l}_6 - \overline{\,l}_5) = 1.0 \pm 0.1$
is a linear combination of renormalized couplings. 
This linear combination is renormalization scale and scheme independent, 
and the value is known at the 10\% level from rare pion decays%
~\cite{Bijnens:2014lea}. 
Comparing with the results of the NJL model calculation%
~\cite{Coppola:2018vkw}, 
we obtain the polarizability
\begin{equation}
\left( \beta_M^{\pi^-} \right)_\text{NJL}
\Big/ \beta_M^{\pi^-}
=
\left( \frac{F_\pi}{M} \right)^2 \frac{17/27}{- I_2(M,\Lambda)}
+ 
\mathcal{O} \left( \frac{m_\pi^2}{4 M^2} \right)
\approx 
- \frac{1}{10}
\label{eq:betaNJL}
,\end{equation}
which disagrees even qualitatively with low-energy QCD.

\section{Axial-Vector and Vector Current Pion Transitions}

The consideration of axial-vector and vector pion matrix elements in magnetic fields is motivated by the pioneering 
lattice calculation of pion weak decay%
~\cite{Bali:2018sey}. 
Among different approaches, 
there is qualitative agreement: 
$\Gamma(\pi^- {\to} \mu \, \overline{\nu}_\mu)$
increases in a magnetic field due to the available energy from Landau quantization, 
and
$\Gamma(\pi^- {\to} e \, \overline{\nu}_e)$
is substantially enhanced due to the magnetic field allowing helicity suppression to be overcome. 
General parameterizations of the required matrix elements in a magnetic field have been given%
~\cite{Coppola:2018ygv}. 
In ChPT, 
we find it convenient to instead parameterize the renormalized current operators, 
from which the matrix elements can readily be calculated%
~\cite{Adhikari:2024vhs}. 
For the effective axial-vector current, 
we write
\begin{equation}
J_A^{\mu-}
=
- 
\left[
F_{\pi^-}^{(A1)} D^\mu
+
i e F_{\pi^-}^{(A2)} F^{\mu \nu} D_\nu
+
e^2 F_{\pi^-}^{(A3)} F^{\mu \nu} F_{\nu \alpha} D^\alpha
\right]
\pi^-
,\end{equation}
which should be compared to the familiar
$B = 0$
current
$J_A^{\mu-} = - F_\pi \partial^\mu \pi^-$. 
The coupling 
$F_{\pi^-}^{(A3)}$
receives non-vanishing contributions in ChPT only starting at 
next-to-next-to-leading order
(NNLO), 
for which we are unable to say anything quantitatively. 
The coupling 
$F_{\pi^-}^{(A2)}$
arises in ChPT at NLO, 
for which there are only tree-level contributions from NLO operators. 
These operators are identical to those in 
Eq.~\eqref{eq:L4}
on account of chiral symmetry, 
and lead to the value
$F_{\pi^-}^{(A2)}
=
\frac{\overline{\ell}}{(4 \pi)^2 F_\pi}$,
where 
$\overline{\ell}$
is precisely the same linear combination of low-energy constants appearing the charged pion magnetic mass
Eq.~\eqref{eq:mpi-}.
This value can be compared to that obtained in the chiral limit of the NJL model study%
~\cite{Coppola:2019uyr}, 
which instead finds
$\big( F_{\pi^{-}}^{(A2)} \big)_\text{NJL} \Big/ F_{\pi^-}^{(A2)} = 2$. 
On the surface, 
this is not much of a discrepancy; 
however, 
it is completely inconsistent with Eq.~\eqref{eq:betaNJL}. 
Chiral symmetry considerations show that both results depend only on one parameter, 
and this cannot be accommodated in the NJL model.

For
$F_{\pi^-}^{(A1)}$, 
which reduces to the pion decay constant at 
$B =0$, 
the ChPT result at NLO is%
~\cite{Tiburzi:2008ma}
\begin{equation}
F_{\pi^-}^{(A1)}
\Big/ F_\pi
=
1 - \frac{e B}{2 (4 \pi F_\pi)^2} \mathcal{I} \left( \frac{m_\pi^2}{eB} \right)
>
1 
\quad 
\Rightarrow
\quad
F_{\pi^-}^{(A1)}
\Big/ F_\pi
=
1 + \frac{1}{3} \left( \frac{eB}{8 \pi m_\pi F_\pi} \right)^2 + \mathcal{O}(B^4)
\label{eq:FA1}
,\end{equation}
where we have indicated that the coupling exceeds 
$F_\pi$
when 
$B \neq 0$
[the function 
$\mathcal{I}(z)$
appears in 
Eq.~\eqref{eq:I}],
and shown this explicitly in the small-field limit, 
for which the dependence is quadratic. 
This amplitude is a quantity for which lattice data exist%
~\cite{Bali:2018sey}, 
however, 
there is not qualitative agreement. 
For small magnetic fields
$e B \lesssim 0.2 \, \text{GeV}^2$, 
lattice values satisfy 
$\big( F_{\pi^-}^{(A1)} \big)_\text{lattice} < F_\pi$. 
To accommodate such behavior, 
the continuum extrapolated results require the empirical form 
$\big( F_{\pi^-}^{(A1)} \big)_\text{lattice} \Big/ F_\pi =  1 - \mathcal{C} \, e B$,
near 
$e B = 0$, 
which is not supported by low-energy QCD, 
Eq.~\eqref{eq:FA1}.

An additional feature of weak pion decay in a magnetic field is the existence of the vector channel, 
which was first recognized in the pioneering lattice work%
~\cite{Bali:2018sey}. 
In ChPT,
the single-pion vector current reads
\begin{equation}
J_V^{\mu -}
= 
- e F_{\pi^-}^{(V)} \widetilde{F}^{\mu \nu} D_\nu \pi^-
,\end{equation}
in terms of the dual field-strength tensor. 
For perturbatively small magnetic fields, 
the source of this contribution is the chiral anomaly. 
This was pointed out in the NJL model study of 
Ref.~\cite{Coppola:2019uyr}, 
for which the underlying loop diagram formed from constituent quarks essentially incorporates the anomaly, 
but was given only very minor consideration.  
In ChPT, 
we obtain the identical result%
~\cite{Adhikari:2024vhs}
$F_{\pi^-}^{(V)} \big/ F_\pi = \frac{2}{(4 \pi F_\pi)^2} = 1.49(3) \, \texttt{GeV}^{-2}$, 
where we have included an uncertainty due to the unknown quark-mass dependent corrections appearing at 
NNLO. 
This value must be contrasted with the lattice results%
~\cite{Bali:2018sey}
\begin{equation}
\frac{
\Big( F_{\pi^-}^{(V)} (B{=}0) \Big)_\text{lattice}
}
{F_\pi}
=
\begin{cases}
1.2(3) \, \texttt{GeV}^2 & \text{Wilson quarks, quenched, } m_\pi = 415 \, \texttt{MeV}
\\
0.8(2) \, \texttt{GeV}^2 & \text{Staggered quarks, fully dynamical, } m_\pi = 135 \, \texttt{MeV}
\end{cases}
,\end{equation}
obtained from two different fermion discretizations. 
Taken together, 
the lattice results are consistent with each other,
however, 
the staggered result has tension with the requirement of the axial anomaly.%
\footnote{
Note well that the fourth-root trick implemented for the staggered fermion determinant does \emph{not} affect this result. 
Even with fully quenched gauge configurations, 
the vector-current matrix element 
$\big\langle 0 \, \big| \, \overline u \gamma_\mu d \, \big| \, \pi^-(p_\parallel) \big\rangle_B$
provides a direct measure of the chiral anomaly using the valence up and down quarks. 
The additional vector current is provided by the magnetic field at linear order, 
and includes the requisite flavor-singlet component. 
Chiral corrections will be sensitive to the quark sea, 
but these do not affect the chiral-limit prediction. 
}

\section{Summary}

We provide a selective overview of constraints on QCD in magnetic fields arising from ChPT. 
Masses of pions and pion transition matrix elements of axial-vector and vector currents are used to 
illustrate low-energy theorems for QCD in a magnetic field.
We point to a few discrepancies with models and tension that exists with lattice results. 
Space does not permit a complete discussion of finite-volume effects related to the quantization of uniform magnetic fields on a torus%
~\cite{Adhikari:2023fdl}. 
The ingenious finite-volume method used to access the magnetization through the pressure anisotropy at fixed magnetic flux%
~\cite{Bali:2013esa}, 
for example,
is potentially susceptible to sizable volume effects.

Looking beyond the topics covered here, 
ChPT will continue to provide constraints for low-energy QCD in extreme environments, 
such as in electric fields and in rotating frames. 
While hadronic models can be adapted to these scenarios, 
low-energy theorems will help assess whether models are trustworthy.
These examples are additionally significant because lattice QCD suffers a sign problem for either of them, 
making rigorous predictions especially challenging. 
Methods, 
such as analytic continuation
($E \to i E$ and $\Omega \to i \Omega$), 
will profit from the anchors provided by 
ChPT.



\begin{thebibliography}{99}

\bibitem{Kong:2022cbk}
L.~D.~Kong \textit{et al.}
Astrophys. J. Lett. \textbf{933}, L3 (2022),
arXiv:2206.04283 [astro-ph.HE].

\bibitem{Skokov:2009qp}
V.~Skokov, A.~Y.~Illarionov and V.~Toneev,
Int. J. Mod. Phys. A \textbf{24}, 5925 (2009),
arXiv:0907.1396 [nucl-th].

\bibitem{STAR:2023jdd}
M.~I.~Abdulhamid \textit{et al.} [STAR],
Phys. Rev. X \textbf{14}, 011028 (2024),
arXiv:2304.03430 [nucl-ex].

\bibitem{Gasser:1983yg}
J.~Gasser and H.~Leutwyler,
Annals Phys. \textbf{158}, 142 (1984).

\bibitem{Tiburzi:2008ma}
B.~C.~Tiburzi,
Nucl. Phys. A \textbf{814}, 74 (2008),
arXiv:0808.3965 [hep-ph].

\bibitem{Coppola:2018vkw}
M.~Coppola, D.~G\'omez Dumm and N.~N.~Scoccola,
Phys. Lett. B \textbf{782}, 155 (2018),
arXiv:1802.08041 [hep-ph].

\bibitem{Bijnens:2014lea}
J.~Bijnens and G.~Ecker,
Ann. Rev. Nucl. Part. Sci. \textbf{64}, 149 (2014),
arXiv:1405.6488 [hep-ph].

\bibitem{Bali:2018sey}
G.~S.~Bali, B.~B.~Brandt, G.~Endr\H{o}di and B.~Gl\"a\ss{}le,
Phys. Rev. Lett. \textbf{121}, 072001 (2018),
arXiv:1805.10971 [hep-lat].

\bibitem{Coppola:2018ygv}
M.~Coppola, D.~G\'omez Dumm, S.~Noguera and N.~N.~Scoccola,
Phys. Rev. D \textbf{99}, 054031 (2019),
arXiv:1810.08110 [hep-ph].

\bibitem{Adhikari:2024vhs}
P.~Adhikari and B.~C.~Tiburzi,
arXiv:2406.00818 [hep-ph].

\bibitem{Coppola:2019uyr}
M.~Coppola, D.~G\'omez Dumm, S.~Noguera and N.~N.~Scoccola,
Phys. Rev. D \textbf{100}, 054014 (2019),
arXiv:1907.05840 [hep-ph].

\bibitem{Adhikari:2023fdl}
P.~Adhikari and B.~C.~Tiburzi,
Phys. Rev. D \textbf{107}, 094504 (2023),
arXiv:2302.09179 [hep-lat].

\bibitem{Bali:2013esa}
G.~S.~Bali, F.~Bruckmann, G.~Endr\H{o}di, F.~Gruber and A.~Sch\"afer,
JHEP \textbf{04}, 130 (2013),
arXiv:1303.1328 [hep-lat].

\end{thebibliography}
\end{document}